\documentclass[12pt,preprint]{aastex}

\slugcomment{.}

\shorttitle{}
\shortauthors{Lieu \& Mittaz}

\begin{document}

\title{On the absence of gravitational lensing of the
cosmic microwave background}

\author{Richard Lieu\altaffilmark{1} \& Jonathan P.D. Mittaz\altaffilmark{1}}

\affil{Department of Physics, University of Alabama at Huntsville,
    Huntsville, AL 35899}

\begin{abstract}
The magnification of distant sources by mass clumps at lower 
($z \leq 1$) redshifts is calculated analytically.  The clumps are
initially assumed to be galaxy group isothermal          
spheres with properties inferred from an extensive survey.                
The average                                 
effect, which includes strong lensing,
is exactly counteracted by the beam divergence
in between clumps (more precisely,
the average reciprocal magnification cancels the inverse Dyer-Roeder
demagnification). 
This conclusion
is independent of the matter density function within each clump,
and remains valid for arbitrary values of $\Omega_m$
and $\Omega_{\Lambda}$.
When tested against the
cosmic microwave background data, a rather large lensing
induced {\it dispersion} in the angular size of the
primary acoustic peaks of the TT power spectrum
is inconsistent with WMAP observations.  The situation is
unchanged by the use of NFW profiles for the density distribution
of groups, which led in fact to slightly larger fluctuations.  Finally, our
formulae are applied to an ensemble of NFW mass clumps or
isothermal spheres having the
parameters of galaxy {\it clusters}.  The acoustic peak size dispersion
remains unobservably large, and is also excluded by WMAP.  For
galaxy groups, two
possible ways of reconciling with the data are proposed,
both exploiting maximally the uncertainties in our knowledge of group 
properties.  
The same escape routes are not available in the case of clusters, however,     
because their properties are well understood.  Here we have
a more robust conclusion: neither the NFW profile nor
isothermal sphere profiles are an accurate
description of clusters, or important elements of physics responsible
for shaping zero curvature space are missing from the standard
cosmological model.  When all the effects
are accrued, it is difficult to understand how WMAP could reveal
no evidence whatsoever of lensing by groups and clusters.
\end{abstract}

\vspace{3mm}

\noindent
{\bf 1. Introduction - observed statistics of galaxy groups}

Light as it propagates through the near Universe will encounter
inhomogeneities.  The motivation which initiated the present paper is to
investigate the
role played by groups of galaxies in the phenomenon of lensing and global
curvature, by
using (perhaps for the first time) real observational data.
Recently, large databases on groups have become available,
including in particular an ESO survey (Ramella et al 2002), from which some
general properties of 1,168 groups can be inferred.  Specifically concerning
the lensing performance of the groups, one needs to know their
mass and velocity dispersion.  These are shown in Figure 1.
We then found the mean virial mass per group to be
\begin{equation}
\overline{M}_{{\rm group}} \approx 1.15 \times 10^{14}
M_\odot.  
\end{equation}
The datum can then be used to estimate the value of $\Omega_{{\rm group}}$, 
since
these 1,168 systems were identified within a surveyed volume of 0.0075 Gpc$^3$
(comoving volume of a pie subtending 4.69 sr at the observer and of radius
extending to redshift $z =$ 0.04, in a $h =$ 0.71, $\Omega_m =$ 0.27,
$\Omega_{\Lambda} =$ 0.73 cosmology).  Hence the number density of groups is:
\begin{equation}
n_{{\rm group}} = n_0 =  1.56 \times 10^{-4}~~{\rm Mpc}^{-3},
\end{equation}
and is in fact
a lower limit, because no attempt was made here to correct 
$n_{{\rm group}}$ for
selection effects.  Eqs. (1) and (2)
point to a mass density of
\begin{equation}
\Omega_{{\rm group}} \approx 0.135.
\end{equation}
Given that most galaxies exist in groups, the value is in agreement with the
conclusion of Fukugita (2003) and Fukugita, Hogan, \& Peebles (1998), who found
(after careful mass budget accountancy) that at low redshifts matter amounting
to 50 \% the total density of $\Omega_m \approx$ 0.27 is connected with
galaxies and galactic environments.

\begin{figure}
\centerline{\includegraphics[height=3.0in,angle=0]{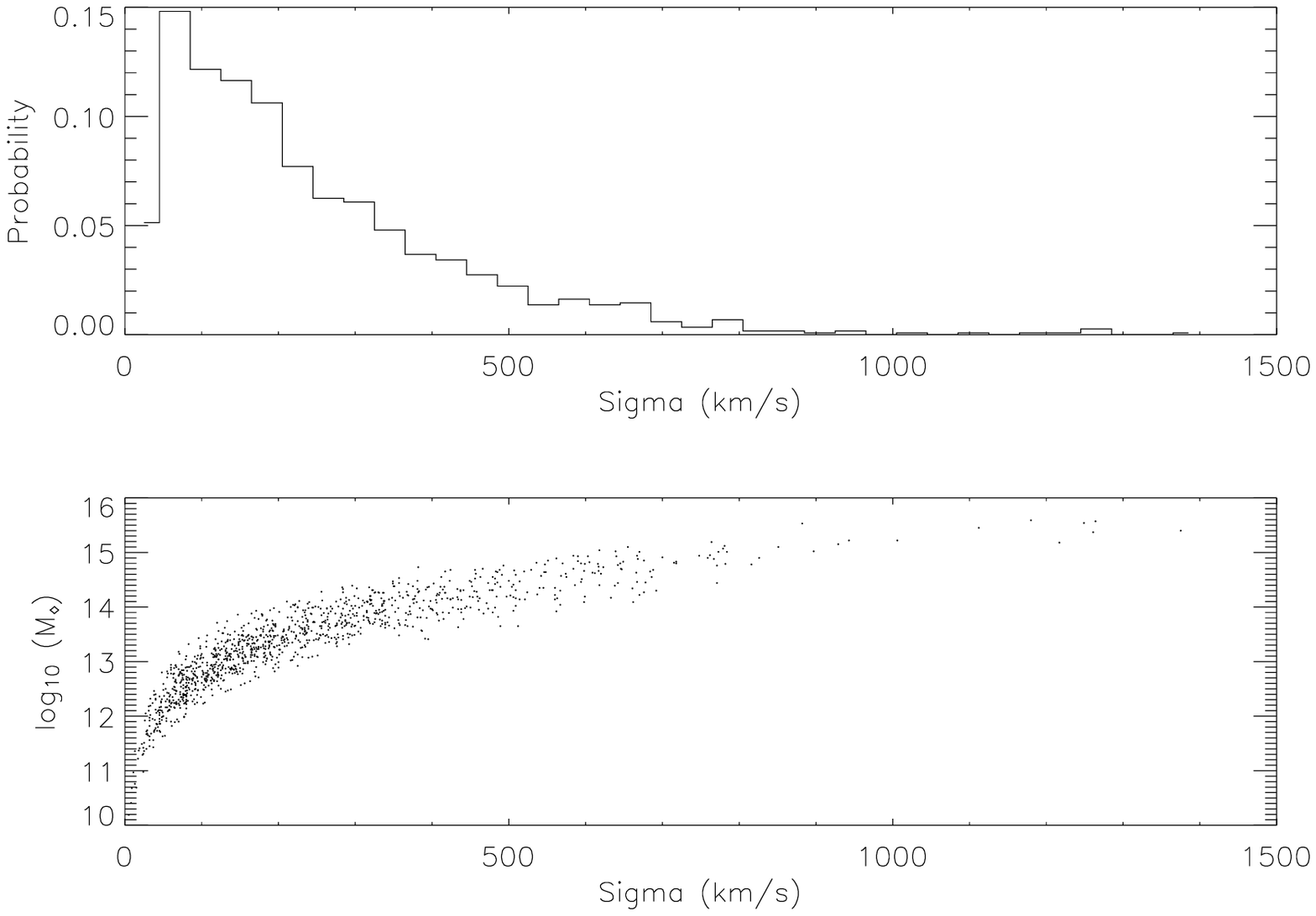}}
\caption{The probability distribution of galaxy group velocity dispersion (top
  panel) and the distribution of group mass against velocity dispersion (bottom
  panel), both derived from the ESO survey (Ramella et al 2002).  These data
  have been used to evaluate the quantity $\sum_{i,j} p_{ij} \sigma_i^4 {\rm
  ln} (R_j/b_{min})$ of Eq. (31).}
\end{figure}

How does such a form of mass concentrations affect the propagation of light?
To answer this question, some information about the matter profile in groups is
necessary.  The limited isothermal sphere
model, wherein the internal matter density falls radially as $1/r^2$ to some
cutoff radius $R$, is sometimes advocated
(Zabludoff \&
Mulchaey 1998, Mulchaey 2000).
In this model the value of $R$ is related to
the total mass $M$ of the group by the equation
\begin{equation}
\frac{GM}{R} = 2 \sigma^2,
\end{equation}
where $\sigma$ is the dispersion velocity of the group.  Since the median
dispersion velocity of the ESO group sample is $\sigma \approx$ 270 km s$^{-1}$
(Ramella et al 2002), Eq. (1) and (3) may be coupled to provide an estimate of
the cutoff radius $\overline{R}$ as $\overline{R} \approx$ 3 Mpc.

\vspace{2mm}

\noindent
{\bf 2. Weak lensing: direct incorporation of galaxy group data}

Concerning the observed breakdown of the Universe's total mass density, viz.
$\Omega_{\Lambda} =$ 0.73 dark energy and $\Omega_m =$ 0.27 matter (Bennett et
al 2003), while at recent epochs ($z \sim$ 1 or less) the 
$\Lambda$ component may remain
smooth, the matter is certainly known to be clumped into mass concentrations,
with galaxy groups forming an important subclass.

The efforts to date on the global (all sky) influence exerted by weak lensing
involve primarily N-body simulations (e.g. Wambsganss et al 1997, Barber 2000),
complemented by some development on theoretical methods (Dalal et al 2003).
There are, however, two areas of neglect: (a) if groups were adequately
represented in previous works, it is doubtful whether direct observational
properties were
involved; (b) analytical formulae are generally lacking, even though they
should be provided to the furthest extent possible - the physics is always much
clearer through this approach.

\vspace{2mm}

\noindent
{\bf 3. The mean convergence for nearby and very distant sources}

Let the origin of the coordinate system be at the observer, who receives a
light signal at world time $\tau_o$.   Let symbols
like $x$ and $y$ denote
the homogeneous Friedmann-Robertson-Walker (FRW) coordinate
distance in the limit when space curvature has the negligible value measured by
the microwave background observations (with WMAP being the latest, Bennett et
al 2003).
Suppose the light was emitted by a source
at coordinates $(x_s,y_s,0)$ and passed through {\it en route} at world time
$\tau$ a galaxy group with its center at position $x$ (i.e. $c \tau=c
\tau_o-x$).  If the group is an isothermal sphere, its potential function will
have the form
\begin{equation}
\Phi(r) = -\frac{GM}{R} \left[1 - {\rm ln}\left(\frac{r}{R}\right)\right].
\end{equation}
Further, assume that the middle ray of a pencil beam skirts by the center of
the sphere at physical distance $b$ ($< R$), equivalent to coordinate
distance $y=b/a(\tau)$, where $a(\tau)$ is the Hubble expansion parameter
at time $\tau$.  The light is deflected inwards by an angle
\begin{equation} 
\psi = \frac{4GM}{R}\left[\arccos\left(\frac{b}{R}\right)+
\frac{R-\sqrt{R^2- b^2}}{b}\right],
\end{equation}
where in Eq. (6) and the
rest of this work the speed of light {\it in vacuo} is set to unity.
Note that
for $b \ll R$, $\psi$ reduces to the familiar constant
$\psi = 2\pi GM/R = 4\pi \sigma^2$.

As a result of the lens, the source appears to us to be at position
$(x_s,y'_s,0)$, where
\begin{equation}
y'_s=\frac{y x_s}{x}.
\end{equation}  
The geometry of the situation requires that
\begin{equation}
y_s=y'_s-\psi (x_s-x).
\end{equation}
The apparent size of the source in the radial direction is affected by an
amount in accordance with $dy'_s  - dy_s = (x_s - x) d\psi$.
The apparent size in the azimuthal direction,
however, is larger than the true size by the ratio $y'_s/y_s$.  Hence the
effect of the scattering is to increase the {\it average} angular size $\theta$
of the source by the fractional amount
\begin{equation}
\eta \equiv\frac{\theta'-\theta}{\theta}
\end{equation}
which by virtue of Eqs. (7) and (8) may be written as
 \begin{equation}
\eta = \frac{(x_s-x)x}{2(1+z)x_s} \left(\frac{\psi}{b} +
\frac{d\psi}{db} \right).
 \end{equation}
where in Eq. (10) and the rest of this paper the expansion 
parameter at epoch $\tau_0$ is set at $a_0 = 1$, and $z \equiv z(x)$
is the redshift of the lens.
This leads to a fractional increase in the energy flux carried by the beam, by
the amount $2 \eta$.

The function $z(x)$ may be obtained by inverting the equation
\begin{equation}
x=c(\tau_0-\tau)=\frac{1}{H_0}\int_0^z\frac{dz'}{E(z')},
\end{equation}
with
\begin{equation}
E(z)=\frac{H(\tau)}{H_0}=[\Omega_m(1+z)^3 + (1 - \Omega_m)]^{1/2}.
\end{equation}
The result is:
\begin{equation}
z(x)=H_0 x +\frac{3}{4}\Omega_m H_0^2 x^2
+\cdots,
\end{equation}
i.e. an expansion of $z(x)$ as a power series in $x$.

In reality the light beam intercepts a random and homogeneous distribution of
galaxy groups along the way.  To work out the total convergence under the
scenario of infrequent multiple scattering, we first write down the probability
of the arriving light having encountered a group at position between $x$ and
$x+dx$, and impact parameter\footnote{Strictly speaking $b$ should be
replaced by the distance of closest approach of the original undeflected ray,
but in the weak lensing limit where $\psi x \ll R$ this distinction is
unimportant.}  between $b$ and $b+db$, as $ndV$.  Here $n$ and
$dV$ are respectively the number density of groups and a cylindrical volume
element, both for the epoch $\tau$, i.e.
 \begin{equation}
ndV =
n(\tau) a(\tau) dx~\times~2\pi b db=2\pi n_0 (1+z)^2\,dx\,b
db.
 \end{equation}
where $n(\tau) \equiv n(z)$ has the functional form $n(z) = n_0 (1+z)^3$ due to
Hubble expansion.  In Eq. (14) the properties of the galaxy groups were assumed
not to change with redshift.  The validity of this statement is restricted to
$z$ being beneath some maximum redshift.  Above the maximum, the evolution of
groups must be taken into account.  This limiting value of $z$ seems to be at
least $z \approx$ 0.5 from direct observation of groups (Jones et al 2002).
From dynamical time considerations, a group cannot evolve significantly within
a timescale $\leq \overline{R}/\sigma \approx$ 3 $\times$ 10$^{17}$ s,
corresponding to a redshift exceeding $z =$ 1.5.

We proceed to calculate the expectation value  of $\eta$, which is given by
\begin{equation}
\langle\eta\rangle= \sum \eta n \delta V = \int \eta n dV
\end{equation}
with the integration performed over an entire cylinder wherein lenses are
present.  By means of Eqs. (6), (10), (14), and (15), we performed
the integration over $b$ to arrive at
\begin{equation}
\langle\eta\rangle=2\pi^2 GM n_0 \int_0^{x_f}dx\,
[1+z(x)]\frac{(x_s-x)x}{x_s},
\end{equation}
where the assumption is for the furthest lens to have 
a present epoch distance of $a_0 x_f = x_f$
(beyond $x=x_f$
the Universe is too smooth to accomodate galaxy groups as we know
them today), and that $b_{\mathrm{min}}\ll R$.  In obtaining Eq. (16) use
was also made of the definite integral
 $$
\frac{1}{R} \int_0^R {\rm arccos} \left(\frac{b}{R} \right) 
d b = 1
 $$

In Eq. (16) we incorporate the contribution to $\langle\eta\rangle$ from all
types of (galaxy group) mass spheres, each having its own mass $M$ and number
density $n_0$, with $\sum n_0 M = \rho_c \Omega_{{\rm groups}}$, where
\begin{equation}
\rho_c = \frac{3 H_o^2}{8 \pi G}
\end{equation}
is the critical density.  We can recast the expression for $\langle\eta\rangle$
as
\begin{equation}
\langle\eta\rangle= \frac{3}{2} \Omega_{{\rm groups}} 
H_0^2 \int_0^{x_f}dx\,
[1+z(x)]\frac{(x_s-x)x}{x_s},
\end{equation}
which is the mean angular magnification of any small patch of sky randomly
located at coordinate position $(x_s,y_s,0)$.  There are two limiting cases
when Eq. (18) simplifies.  The first is $x_s=x_f$, corresponding to sources
embedded within the clumpy environment of the near Universe (e.g. Type 1a
supernovae). In this case
\begin{equation}
\langle\eta\rangle= \frac{1}{4} \Omega_{{\rm groups}}
H_0^2 x_s^2 \left( 1 + \frac{1}{2} H_0 x_s
+ \frac{9}{20} \Omega_m H_0^2 x_s^2 + \cdots
\right)
\end{equation}
The 2nd is $x_s \gg x_f$, corresponding to very distant sources which emitted
light at a time when the Universe was smooth, the cosmic microwave background
(CMB).  Here we have
\begin{equation}
\langle\eta\rangle= \frac{3}{4} \Omega_{{\rm groups}} 
H_0^2 x_f^2 \left( 1 + \frac{2}{3} H_0 x_f
+ \frac{3}{8} \Omega_m H_0^2 x_f^2 + \cdots \right).
\end{equation}
Note the absence of any $x_s$ dependence in Eq. (20).  If corrections due to
the finiteness of $x_f/x_s$ are desired, we remark that the lowest order of
such terms equals an additional $-2x_f/(3x_s)$ within the last pair of
parentheses.

\vspace{2mm}

{\bf 4. The exact cancellation between the lensing effect of
isothermal spheres and the Dyer-Roeder demagnification}

A very interesting result which emerges from the analysis thus far 
concerns the balance between beam convergence by the clumps and
divergence within the subcritical density `voids' between clumps.
Take for instance a supernova source, the weak lensing of which
is described by Eq. (19).  If the `voids' are completely matter-free, i.e.
$\Omega_g = \Omega_m$ in Eq. (19), one would attain maximum
$\langle\eta\rangle$.  Yet, in this same limit, the demagnification $\epsilon$
of the source caused by propagation through the `voids' will be that of
the Dyer-Roeder `empty beam' (Dyer \& Roeder 1972), viz.
 $$
\epsilon=\frac{x'_s - x_s}{x_s}
 $$
where
\begin{equation}
x_s = \frac{1}{H_0} \int_0^{z_s} \frac{dz}{E(z)}~;~
x'_s = \frac{1}{H_0} (1+z_f) \int_0^{z_f} \frac{dz}{(1+z)^2 E(z)},
\end{equation}
and $E(z)$ as defined in Eq. (12).
The remarkable fact is that an expansion of $\epsilon$ in power series
of $x_s$ with the aid of Eq. (13) gives {\it the same series as Eq. (19)
with $\Omega_g = \Omega_m$}.

This development highlights the advantage of an analytical approach.
In an earlier work, Weinberg (1976) considered
an $\Omega_{\Lambda} =0$ Universe where all the matter is 
clumped into point masses, and found that the net magnification is still
controlled by the $\Omega_m$ parameter alone - as if space
remained homogeneous.  The present conclusion reinforces Weinberg in
the more realistic context involving clumps of finite size within a
Universe of arbitrary $\Omega_{\Lambda}$ and $\Omega_m$.

The correspondence between
a 100 \% clumped Universe and
the fully homogeneous Universe, as established above, remains in place
even if the
clumping is not 100 \%.  This was shown in a separate paper (Lieu \&
Mittaz 2004) where we also adopted a more unifying method, using
the Sach's optical equations to handle the two opposing effects under
one formalism.  A toy model on the physics of this section is given
in Appendix A.  The formal treatment of average magnification in
an inhomogeneous Universe is provided by Kibble \& Lieu (2005),
where the diversity of averages appropriate to different
modes of observations and data analysis methods
will be calculated and discussed.

\vspace{2mm}

\noindent
{\bf 5. The standard deviation - convergence fluctuations}

Like the two-point correlation function in galaxy count analysis, the first
step towards an expression for $\delta \eta$ is to let $n_i$ be the number of
groups with their centers lying within a volume $\delta V$ labelled
positionally by an index $i$.  Provided $\delta V$ is sufficiently small that
there is no appreciable chance for the centers of two groups to be both inside
volume $i$, then $n_i =$ 0 or $n_i =$ 1, i.e.
\begin{equation}
n^2_i = n_i
\end{equation}
Moreover, because the distribution of groups in space is a Poisson process, and
the location of each group does not affect that of another, $n_i$ has the
following properties concerning its averages:
\begin{equation}
\langle n_i \rangle = n \delta V,~~
\langle n_i^2 \rangle = n \delta V,~~
\langle n_in_j \rangle =
\langle n_i \rangle \langle n_j \rangle = (n \delta V)^2.
\end{equation}
For each volume there is a corresponding contribution to the convergence.  If
the $i^{th}$ cell is occupied, we will have
\begin{equation}
\eta_i = \frac{(x_s-x_i)x_i}{2(1+z_i)x_s} \left(\frac{\psi_i}{b_i} +
\psi'_i \right)
\end{equation}
as the fractional change in the angular size of the source
due to its light passing by this cell ($\psi' = d\psi/db$).
The total effect is given by a
summation along the light path:
\begin{equation}
\eta = \sum_i n_i\eta_i.
\end{equation}
Taking the average, we have
\begin{equation}
\langle\eta\rangle =
\sum_i \langle n_i \rangle \eta_i
= \sum_i n\delta V \eta_i.
\end{equation}
i.e. one obtains Eq. (15) for $\langle\eta\rangle$.

Continuing towards the variance, we need the average value of
\begin{equation}
\eta^2 = \left(\sum_i n_i\eta_i \right)^2
= \sum_i n_i\eta_i^2 +
\sum_{i\ne j} n_in_j\eta_i\eta_j,
\end{equation}
which is given, after taking into account Eq. (23), by
\begin{equation}
\langle\eta^2\rangle
= \sum_i (n \delta V)\eta_i^2 +
\sum_{i\ne j} (n \delta V)^2\eta_i\eta_j.
\end{equation}
The variance is now computed in accordance with its definition:
\begin{equation}
(\delta \eta)^2=\langle\eta^2\rangle-
\langle\eta\rangle^2
=\sum_i [n \delta V-(n \delta V)^2]\eta_i^2.
\end{equation}
where use was made of Eq. (26) and the fact that the $\langle\eta\rangle^2$
term cancels unless $i=j$.  Finally, we note that for small enough $\delta V$,
$(n \delta V)^2\ll n \delta V$ and may be ignored, so that $(\delta \eta)^2 =
\sum_i n \delta V \eta_i^2$.
If the summation over $i$ is cast
in integral form, we will have
\begin{equation}
(\delta \eta)^2= \int \eta^2 ndV = 8 \pi^3
n_0 \sigma^4 \left[
\ln\left(\frac{R}{b_{\mathrm{min}}}\right) - \frac{8}{\pi^2} \right]
\int_0^{x_f} dx\,
\left[\frac{(x_s-x)x}{x_s}\right]^2,
\end{equation}
where in going towards the last expression use was made of Eqs. (10) and (14)
with the full deflection angle from Eq. (6).

Note that the standard deviation differs from the mean $\langle\eta\rangle$ in
two distinct ways.  First, while $\langle\eta\rangle \propto 
\Omega_{{\rm groups}}$,
$(\delta \eta)^2$ is more complicated, being $\propto n_o \sigma^4$.
Thus $(\delta \eta)^2$ is much more governed by
the specific properties of the type of isothermal spheres in question.  Second,
unlike $\langle\eta\rangle$, the integration towards $(\delta \eta)^2$ can be
performed exactly for all values of $x_f$ and $x_s$.  When this is done, and
the variance for groups of various radii and velocity dispersions are summed,
the result is
\begin{equation}
(\delta \eta)^2= \frac{8 \pi^3}{3} n_0 x_f^3
\left( 1 - \frac{3x_f}{2x_s}
+ \frac{3x_f^2}{5x_s^2} \right)
\sum_{i,j} \left\{ p_{ij} \sigma_i^4 \left[
{\rm ln} \left(\frac{R_j}{b_{{\rm min}}}\right) - \frac{8}{\pi^2}
\right] \right\},
\end{equation}
where $p_{ij}$ is the probability of finding a group with velocity dispersion
$\sigma_i$ and radius $R_j$.

\vspace{2mm}

\noindent
{\bf 6. Application to Type 1a Supernovae - a check against
numerical simulations}

Utilizing the 
full statistical properties of the 1,168 galaxy groups observed during the
ESO survey (Ramella et al 2002, section 1 and Figure 1), the summation in
Eq. (31) was found to have the value
\begin{equation}
\sum_{i,j} \left\{ p_{ij} \sigma_i^4 \left[
{\rm ln} \left(\frac{R_j}{b_{min}}\right) - 
\frac{8}{\pi^2} \right] \right\} =
2.90 \times 10^{-10}
\end{equation}
for\footnote{At such a small minimum impact parameter
the lensing may have become strong, so that readers can legitimately
question whether the present calculation, which is based on the
assumption of weak lensing, still applies.  The answer is yes, and
is explained in section 8 and Appendix B} $b_{min} =$ 10 kpc and
under the convention of unit speed of light.
The equation, together with Eqs. (2) and (3),
permit an evaluation of the quantities $\langle\eta\rangle$ and $\delta \eta$
as given in Eqs. (18) and (30).  

We now apply the results to the astrophysical
phenomenon
of Type 1a supernovae (SN1a), where $x_f = x_s$ (see the comment after
Eq. (18)), both being $\leq$ a few Gpc, and Eq. (18) reduces to Eq. (19).
At redshifts $z_f=z_s =$ 1 and 1.5, the values of ($\langle\eta\rangle,\delta 
\eta$) are respectively (0.031, 0.036) and (0.060,0.055).  Note in each
case $\delta \eta$ is at most $\simeq \langle\eta\rangle$ (the same
applies also to the CMB - see below).
This means in the context of our paper there is no concern over
the probability distribution of $\eta$ becoming
asymmetric (skewed).  In fact the distribution
is truncated
only at the tail end of the left wing: because from
section 4 it was proved that $ \langle\eta\rangle$
exactly equals the
demagnification percentage at the voids,
a negative excursion of
$\eta$ from the mean value of $\langle\eta\rangle$ by one standard deviation
$\langle\eta\rangle - \delta \eta$ does not render the source fainter than
the rigid lower bound set by the
Dyer-Roeder beam.

In fact, our value of
$2 \delta \eta \approx$ 6 \% for the brightness fluctuation at
$z=1$ compares well with that of cosmological N-body simulations
by Barber (2000), who obtained 7.8 \% (this rather large $\delta \eta$ may
well be the reason why Barber (2000) found an average amplification of -3.4 \%
when our analytical calculation gives zero).  We also quote, for completeness,
another result for $\delta \eta$ from an earlier simulation of Wambsganss
et al (1997), whose value was 4 \%.  The fluctuation is nonetheless still
marginal for the purpose of testing against SN1a data, wherein the
brightness dispersion is at the 0.35 magnitude level (Barris et al 2004,
Tonry et al 2003).
 
\vspace{2mm}

\noindent
{\bf 7. Application to the CMB - should space be as flat as observed?}

Observations of the CMB are much more accurate than those of SN1a, and can be
used to conduct a useful test of the standard cosmological model.  For the CMB
at $z_s \approx$ 1000, and assuming that the galaxy groups populate the near
Universe (without significant evolution) to $z = z_f$ we have, from Eq. (18),
$\langle\eta\rangle =$ 0.099 and 0.157 respectively for $z_f =$ 1 and 1.5.
Also, from Eqs. (2), (31), and (32), we find that $\delta \eta =$ 0.093 at $z_f
=$ 1 and 0.133 at $z_f =$ 1.5.

\begin{figure}
\centerline{\includegraphics[height=3.0in,angle=0]{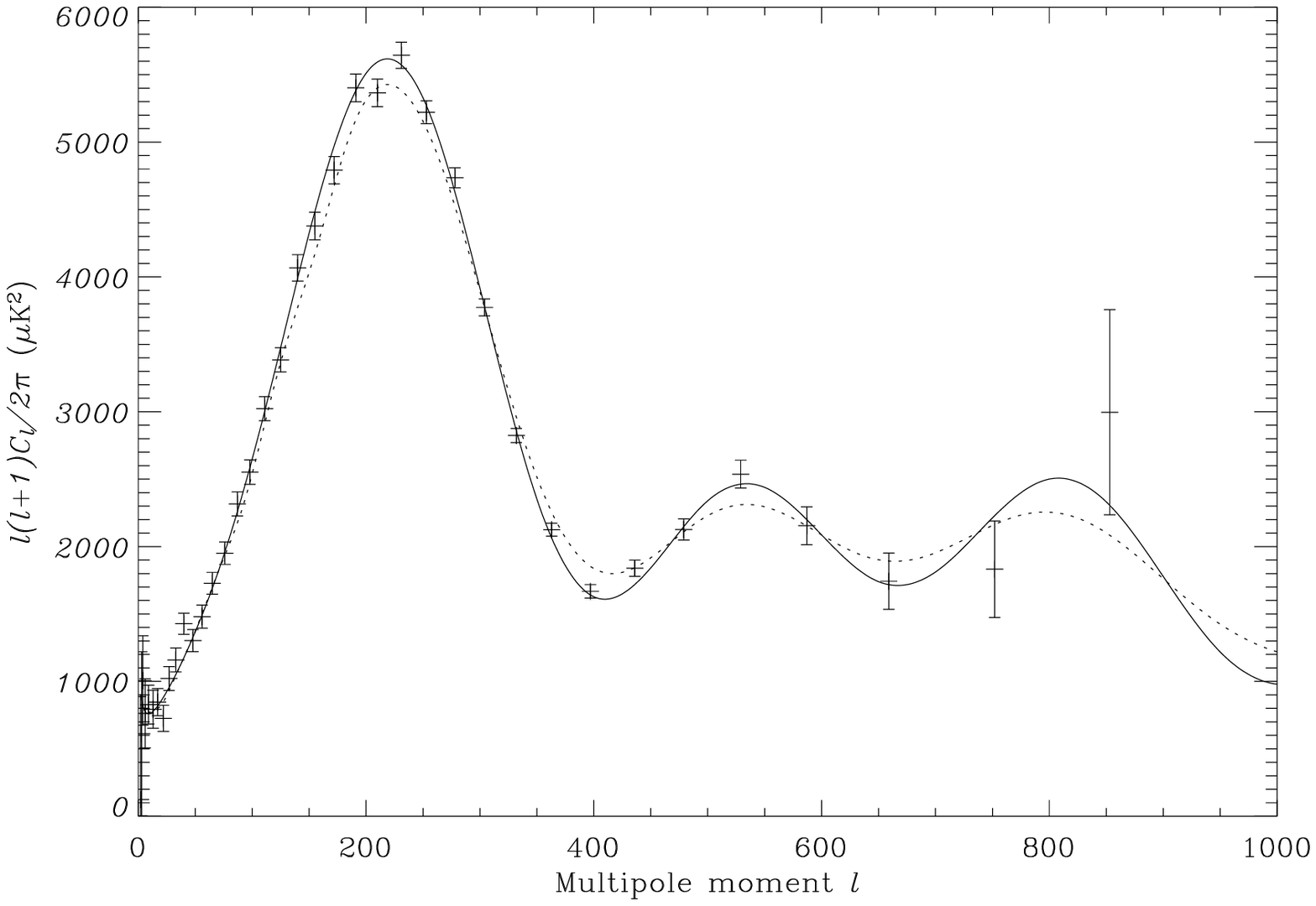}}
\caption{The standard FRW model with $h =$ 0.71, $\Omega_m =$ 0.27, 
$\Omega_{\Lambda} =$
0.73 as applied to the TT power spectrum measured by WMAP
is plotted as a solid line.  If in this model 
50 \%  of the matter within $z \approx$ 1 is clumped into isothermal spheres
with properties given by the ESO survey of galaxy groups, the 
lensing induced convergence
fluctuation ($\delta \eta \approx$ 10 \%) 
will cause a smearing of the spherical harmonics
at and around the primary acoustic peaks by an amount in accordance with
Eqs. (33) and (34).   The resulting theoretical power spectrum is plotted
as a dashed line - it is an unacceptable fit to the data, with 
$\chi^2_{{\rm red}} = 2.42$ for 38 degrees of freedom (null probability
hypothesis $<0.0001$).
}
\end{figure}

In the present work we focus on $\delta \eta$, the dispersion in the angular
size of CMB hot and cold spots.  Previous attempts in estimating the extent of
this effect involved ray tracing codes (e.g. Pfrommer 2003), concerning which
it is not entirely clear what galaxy groups properties were assumed, and how
well they correspond to observations.  The quantity $\delta \eta$ has the
meaning of a standard deviation in the angular size of sources positioned along
independent sightlines that sample the variation in the spatial location of
different sets of groups.  The separation $\alpha$ between such sightlines is
$\sim$ the angular diameter of a typical group, with present physical radius
$\approx$ 3 Mpc (see Eq.(4)), placed midway (in comoving FRW distance scale)
between us and $z=z_f$, i.e.  $\alpha \approx$ 12.4 arcmin for $z_f =$ 1, and
9.4 arcmin for $z_f =$ 1.5.  Hence if a spherical harmonic in the CMB TT
power-spectrum has mean angular size $\theta = \pi/l$ less than or of order
$\alpha$, the dispersion in $\theta$ will simply be
\begin{equation}
\delta \theta 
= \theta \delta \eta~(\theta \leq \alpha,~{\rm coherent~scattering}).
\end{equation}
If on the other hand the structure is large, and it contains $N \sim
\theta^2/\alpha^2$ subregions magnified independently by distinct lens
combinations, then intuitively
\begin{equation}
\delta \theta = \frac{\theta \delta \eta}{
\sqrt{N}}~(\theta > \alpha,~{\rm incoherent~scattering}),
\end{equation}
although Eq. (34) may also be established firmly by a mathematical
proof, which is provided in Appendix C.
In both cases $\delta \theta$ is to be added in quadrature to the intrinsic
dispersion in the structure sizes, which are due to density perturbations at
decoupling.  The dividing line between the two cases, $\alpha \approx$ 10
arcmin, is to be compared with the value of 20 arcmin advocated by Bartelmann
\& Schneider 2001 (BS).  In fact, if the essential part of our Eq. (31), viz.
\begin{equation}
(\delta \eta)^2  \sim \frac{n_0 \sigma^4 L^3}{c^4}
\end{equation}
where $L = x_f$, is converted into the notation of BS, we will make the
substitution $n_0 \rightarrow 1/R^3$ (BS 
assumed $L/R$ lenses per distance $L$), $R \rightarrow 1/k$,
$L \rightarrow w$, and $\sigma^2/c^2 \rightarrow 2\Phi$.
It will then be clear from Eqs. (33) through (35) that $\delta \theta$ is
$\approx$ the deflection angle dispersion $\sigma(\phi)$ of BS for both
coherent and incoherent scattering.  The key difference, however, is that
because for the galaxy groups being considered $\delta \eta \approx$ 0.1, some
five times higher than the corresponding value in BS, one expects significant
broadening of the spherical harmonics.  As can be seen in Figure 2, such a
behavior is inconsistent with the WMAP data.

An immediate check concerns whether the galaxy group properties we derived from
the ESO survey are representative of the truth.  There are two aspects open to
critique.  Firstly, the cutoff radius given in Eq. (4) is determined from the
virial mass $M$ and the dispersion velocity $\sigma$, the former of which
(i.e. $M$) is only an inferred quantity.  It is entirely possible that in
reality we have a mean cutoff radius $\overline{R} \ll$ 3 Mpc (i.e.
the total mass of a typical group is $\ll$ its virial mass), but the total
amount of matter within galaxies and groups constitute an 
$\Omega_{{\rm group}}$
which still satisfies Eq. (3).  This would involve a smaller $M$ and $R$ for
each group, but a larger number density of groups than the value of Eq. (2).
In fact, from an earlier ESO survey Ramella et al (1999)
estimated that if observational selection effects were taken into account
$n_{{\rm group}}$ could reach 4 $\times$ 10$^{-3}$ Mpc$^{-3}$ (see Figure
6c of Ramella et al 1999, where the number density plotted should be multiplied
by $h^3$), i.e. an increase from Eq. (2) by a factor of $\sim$ 25.  To remain
in compliance with Eqs. (3) and (4), then, the cutoff radius $\overline{R}$
(and hence mean angular size $\alpha$) of the groups must both decrease by the
same factor, to $\overline{R} \approx$ 120 kpc.  Returning to Eqs. (31) and
(32), we see that $\delta \eta$ becomes somewhat higher, but the real
difference comes from the $\sqrt{N}$
reduction factor for incoherent scattering at a given CMB spot size, Eq. (34),
which is now 25 times more severe.  As a result of these modifications, the net
broadening effect on the primary acoustic peaks is kept drastically in check -
the theoretical TT power spectrum is no longer distinguishable from that of the
standard model (the solid line of Figure 2).

The second way of resolving the observational conflict is to question the
legitimacy in our assumption that all the observed groups are virialized
systems with the density profile of an isothermal sphere.  Strictly speaking,
the only relatively secure candidates are those 61 X-ray emitting groups in the
ESO survey (Ramella et al 2002), for reasons already explained in section 1.
Since these objects constitute $\approx$ 5 \% of the total sample, it could be
argued that the actual value of $\delta \eta$ should involve a reduced number
density, $n_0 \rightarrow 0.05 n_0$, in which case $\delta \eta$ would become
$\approx$ 2 \%, and one recovers the minimal distortion advocated by BS, with a
resulting TT power spectrum that differs negligibly from the solid line of
Figure 2.  With this reduction of $\delta \eta$, however, the SN1a brightness
dispersion estimate becomes 0.016, closer now to the result of Wambsganss et al
(1997) than Barber (2000).  One question worthy of consideration is, even if
most of the groups are not isothermal spheres, they should still possess some
intra-group gravitational potential, so that the present undertaking of
ignoring altogether their influence on light may not be justified.  Effectively
the problem is resolved under this scenario by assuming that groups
are nothing more than a collection of individual galaxies which act as a large
number of tiny incoherent lenses, thereby minimizing the size dispersion of
large light sources.

\vspace{2mm}

\noindent
{\bf 8. Generalization of sections 3 -- 5 to clumps of arbitrary
internal density profiles}

The development of sections 3 -- 5 may readily be extended to
the case of mass clumps with an arbitrary internal density distribution,
with the deflection angle $\psi (b)$ of Eq. (6) having the general form
\begin{equation}
\psi(b) = 2\int_{-\infty}^{\infty} \frac{Gm(r)b}{r^3} dx =
4\int_0^{\pi/2} \frac{Gm(r)}{b} {\rm cos} \alpha d\alpha,
 \end{equation}
where $r = \sqrt{x^2+b^2} = b {\rm sec}\alpha$, and the mass $m(r)$
within radius $r$ can be an arbitrary function of $r$.    In terms of
the density $\rho(r)$ at $r$, where
 \begin{equation}
\frac{dm(r)}{dr} = 4\pi r^2 \rho(r),
 \end{equation}
the fractional weak lensing angular magnification reads
 \begin{equation}
\eta=\frac{2G x_l(x_s-x_l)}{(1+z_l)x_s} \int_b^\infty
\frac{4\pi r \rho(r) dr}{\sqrt{r^2 - b^2}}.
 \end{equation}
By integrating over the probability element for randomly located
clumps, $dP = ndV$ of Eq. (14), and taking account of the fact that
\begin{equation}
\int_0^{\infty} bdb \int_b^{\infty}
\frac{4\pi r \rho(r)dr}{\sqrt{r^2 - b^2}} =M,
\end{equation}
the total mass of each clump, one again arrives  at Eq. (18) for
the average value of $\eta$.  Thus the conclusion in section 4 of
zero  net magnification relative
to that in a homogeneous Universe of the same mean density is valid
for lensing clumps with any profile $\rho(r)$, and is not
contingent upon the state of inhomogeneity of the Universe.
Moreover, in Appendix B we shall prove that, as long as
the meaning of the average is appropriately defined, both this
conclusion and Eq. (18) for $\langle\eta\rangle$ remain unchanged
even when the light rays are allowed to pass through clumps at
sufficiently small impact parameters where strong lensing effects
must also be included.

In fact,
the most general treatment of the problem is to be found in Kibble \&
Lieu (2005), who showed using a more powerful (vierbein) formalism that
even under a broader range of circumstances than those enumerated
above (a) the Universe may still
be regarded as homogeneous concerning the mean properties of its
propagating light, {\it and}
(b) Eq. (18)
is the correct expression for $\langle\eta\rangle$.
One advantage of Kibble \& Lieu (2005)
lies in its utilization of the
expansion $\theta$ of a small ray bundle, which is an additive
quantity irrespective of the strength of the lensing.

Since the averaging
procedure $\langle\eta\rangle = \int \eta dP$ is
applicable to  the full range of $b$, large and small, it follows from
the derivation of section 5 that the same
statement may be made about the convergence fluctuation, viz. it too is 
always given
by the expression $(\delta \eta)^2 = \int \eta^2 dP =
\int \eta^2 ndV$.  Thus, from Eqs. (38)
and (14)
we deduce that
\begin{equation}
(\delta \eta)^2 = \int \eta^2 ndV =
8 \pi n_0 G^2 \int_0^{x_f}dx_l\,
\left[\frac{(x_s - x_l)x_l}{x_s} \right]^2
\int_0^{\infty} bdb \left[\int_b^{\infty}
\frac{4\pi r \rho(r)dr}{\sqrt{r^2 - b^2}} \right]^2,
\end{equation}
for spherical clumps having any $\rho(r)$.

\noindent
{\bf 9.  NFW profiles}

With the availability of Eq. (40) 
the way is paved for investigating convergence
fluctuation due to the intra-group matter distribution having
a profile $\rho(r)$ which differs from the isothermal sphere.
In particular,
attention is devoted to the increasingly employed Navarro-Frenk-White
or, NFW, model (Dubinski \& Carlberg 1991; Navarro et al 1995, 1996, 1997),
which involves a function of the form
\begin{equation}
\rho(r) = \frac{\delta_c \rho_c r_s^3}{r(r+r_s)^2}
\end{equation}
between $r=0$ and $r=R$, where $\delta_c$ and $r_s$ are respectively known
as the overdensity factor and scale radius, $\rho_c$ is as given
in Eq. (17), 
and $R$ is the virial radius, related to the 
r.m.s dispersion velocity $\sigma$ by
\begin{equation}
R = \frac{\sqrt{3}}{10} \frac{\sigma}{H_0 E(z)},
\end{equation}
with $E(z)$ as in equation (12) and $\sigma$ as a velocity measured
in the frame of the cluster, i.e. $\sigma \sim (1+z)$.  For
$z \leq$ 1.5, $(1+z)/E(z)$ is a constant.  Hence $R$ may also
be regarded as constant, apart from intrinsic evolution
effects which we already
argued is without the support of evidence in the case of groups.
Moreover, the scale radius $r_s$ is $\approx$ 26 \% of the virial radius, or
\begin{equation}
\frac{R}{r_s} = 3.846 = c
\end{equation}
(see Carlberg et al 1997),
and the overdensity $\delta_c$ depends on the ratio $c$  via the equation
\begin{equation}
\delta_c =
\frac{200}{3} \frac{c^3}{{\rm ln} (1+c) - \frac{c}{1+c}}.
\end{equation}
Since we are considering the lensing effects of nearby groups, the
modification to $R$ by the function $E(z)$ is ignored.

From Eqs. (40) and (41) one arrives at
\begin{equation}
(\delta \eta)^2
= \frac{64 \pi^3 n_0 x_f^3}{3} \left[\left(\frac{GM_0}{b_{{\rm min}}}\right)^2 -2 \left(\frac{GM_0}{R}\right)^2 {\rm ln} \left(\frac{R}{b_{{\rm min}}}
\right) \right] \left( 1 - \frac{3x_f}{2x_s}
+ \frac{3x_f^2}{5x_s^2} \right),
\end{equation}
valid for any
$b_{{\rm min}}$ except $b_{{\rm min}} \ll r_s$,
where the quantity $M_0$ is given by
\begin{equation}
M_0 = \delta_c \rho_c r_s^3.
\end{equation}
For galaxy groups with $\sigma \approx$
270 km s$^{-1}$ we obtain, from Eqs. (42) to (44), the
following NFW model parameters: $R =$ 668 kpc, $r_s =$ 174 kpc, and
$\delta_c =$ 4.83 $\times$ 10$^4$.
Substituting these values into Eq. (45), we
computed $\delta \eta$ assuming a group density $n_0$ such that
$n_0 M = \Omega_{{\rm g}}$, 
where $M$ is the total mass within the virial radius $R$,
and $\Omega_{{\rm g}}$ is still given by Eq. (3).
This involved assigning $n_0$ a value $\approx$ three times higher than
that of Eq. (2).  Moreover, by 
setting\footnote{We
conservatively avoid setting $b_{{\rm min}}$ too low, because
the performance of NFW profiles in the very inner
parts of galaxy groups and clusters
is controversial.}
$b_{{\rm min}} \approx$ 
100 kpc,
we then find that $\delta \eta$ is about 1.5 times higher than its
corresponding value
as determined by Eqs. (2),
(31) and (32) for isothermal sphere density profiles, when the 
same distances $x_f$ and $x_s$ apply to both cases.  Thus one can safely
declare that the conclusions of sections 6 and 7 also hold for
NFW density profiles.

\vspace{2mm}

\noindent
{\bf 10. The absence of convergence fluctuation in the CMB due to
clusters of galaxies}

With all the physical and mathematical prerequisites in place, we are
ready for the clincher test, which utilizes  clusters of
galaxies -  extensively observed
systems with well determined properties - as gravitational lenses.
Nearby clusters have a
number density at the present epoch (Bahcall 1988) of
\begin{equation}
n_0 \approx 10^{-5} h^3~{\rm Mpc}^{-3}
\end{equation}
at $h=0.71$, and a mean velocity dispersion
in the cluster frame of
(Struble \& Rood 1991) of
\begin{equation}
\sigma \approx 1000 (1+z)~{\rm km~s}^{-1}.
\end{equation}
The latter corresponds, by the low $z$ version of
Eq. (42), to a mean virial radius of
\begin{equation}
R \approx 2.12~{\rm Mpc},
\end{equation}
and hence a mean virial mass in the range
\begin{equation}
M = \frac{800\pi}{3} \rho_c R^3 \approx 1.12 \times 10^{15} M_\odot.
\end{equation}

Moreover, there is ample evidence that clusters do not evolve
significantly put to $z_f \sim$ 1 (Hashimoto et al 2002, 2004;
Maughan et al 2003; Younger, Bahcall, \& Bode 2005), so that once
again for our present purposes $n_0$ and $R$ may be treated as constants.
 
Combining equations (40) through (44) with equations (47) and (49),
followed by a numerical integration of equation (40) with
$b_{{\rm min}} =$ 100 kpc, $x_s \approx$ 14.02 Gpc to the CMB, and
$x_f \approx$ 4.41 Gpc to the farthest lens (set at $z_f =$ 1.5), one
obtains $\delta \eta \approx$ 10 \%.  Given that nearby clusters
have comparable angular sizes as the galaxy groups (both having $R \approx$
2 -- 3 Mpc), i.e. the effect of incoherent lensing  is already calculated
in section 7,
the implication of this result on the TT primary
acoustic peaks is again an expected smearing of the gaussians as
depicted in Fig. 2 (ignoring a slight skewness
in the convolution), which is contradicted by observations.
 
Alternatively one could return to the isothermal sphere model, with
$\rho(r) = M/(4\pi r^2 R)$ cutting off at $r=R$, and the values of
$R$ and $M$ as given by equations (49) and (50).  Then, with the same
source-lens distances as in the case of the NFW profile, equation (40)
gives also $\delta \eta \approx$ 10 \% at
$b_{{\rm min}} =$ 100 kpc after a numerical integration, i.e.
our conclusion of a conflict between prediction and reality remains.

\vspace{2mm}

\noindent
{\bf 11 Summary and conclusion}

The all-sky convergence fluctuation of light by galaxy groups,
with properties inferred from an extensive survey of groups,
is computed.  When applied to Type 1a supernovae brightness - a situation
where the effects are generally hard to measure
- the results were found to be in
broad agreement with the predictions
of earlier authors.  When considering the CMB, the
consequence of a sizable dispersion in the angular diameter of the
temperature fluctuations is more serious, because it does not correspond to the
observational reality represented by the latest WMAP data.

There are at least two ways of overcoming the difficulty, both of which point
to the possibility that in reality galaxy groups have by far
not yet developed
to their virial masses and virialized potential.
The first one advocates the
existence of many more groups than the directly detected ones
upon which the analysis in this paper was based.
The total number density remains in accordance with
the published correction for selection effects (Ramella et al 1999), but the
mean mass per group is substantially reduced.  The outcome is a great deal more
incoherence in the lensing of the primary acoustic peak structures, and any
additional dispersion in a spherical harmonic 
$\ell$ becomes unobservable.  In the
second resolution we questioned 
the actual fraction of the groups within the ESO
survey sample which have fully formed
isothermal sphere or NFW profiles.  If this is
estimated by counting only the X-ray emitting groups, the number of such groups
is so small that their effect on the CMB will again be beneath the
margin of detectability.

Finally, our predicted  all-sky variation in the size of the CMB
acoustic peaks based upon the lensing effect of clusters of galaxies
was compared with WMAP data.  This raises the much more serious question
on why
convergence fluctuations in the size of the second CMB
acoustic peak due to clusters are absent.  The 
`escape route' arguments of the previous 
paragraph, which may have worked in the case of galaxy groups,
are no longer so compelling in the present context,
simply because
rich clusters are
very well studied systems.
Nevertheless, one could still contemplate
a less profound consequence, viz. perhaps the NFW
and isothermal sphere profiles are
not a good description of clusters after all.  Given, however, that the range
of impact parameters we used to derive the value of $\delta \eta$ is
optimal for the performance of the profiles, this does not seem to
be a sensible way out.  We therefore end with the startling conclusion
that the large scale curvature of space may not
entirely be an initial value problem related to inflation.  The
absence of gravitational lensing of the CMB point to the possibility
that even effects on light caused by wrinkles in the space of
the late (nearby) Universe have been compensated for, beyond some
distance scale, by a mechanism which maintains a flat geometry over
such scales.

Authors are indebted to T.W.B. Kibble for his independent re-deriving
and cross-checking of
the mathematical formulae in the paper.

\newpage

\noindent
{\bf Appendix A - Heuristic model  to illustrate the coupling between
the influence of clumps and voids on light propagation}

We provide a simple way of gaining an insight into why the 
geometry of space as revealed by light is, to the lowest order of
approximation, independent of the state of inhomogeneity of the
incipient matter.   For the purpose it is only necessary to use
non-expanding Euclidean space as starting point, i.e. let space be
uniformly flat and empty, so that the propagation of light is governed
by null geodesics in
a Minkowski background metric.  Clumps may be introduced without
changing the average properties, by placing at random locations
spheres of total mass $M$ and internal density profiles $\rho(r)$
extending to a
radius $R$, with the matter for each sphere being drawn
evenly
from a concentric cavity of radius $R_1 \gg R$.  The picture is then rather
akin to the `swiss cheese' model.  Rays passing at impact parameters
$b > R_1$ behave as if the relevant clump does not exist.  At
$R < b \leq R_1$ a ray is deflected inwards by the angle $\psi_0 =
4GM/b$ because of the clump, and outwards by the angle
 $$
\psi_1 = -\frac{4GM}{b} \left[1-
\frac{(R_1^2- b^2)^{\frac{3}{2}}}{R_1^3} \right] \eqno(A1)
 $$
because of negative mass distribution in the void.
The net inward deflection $\psi = \psi_0 - \psi_1$ is then given by
$$
\psi = \frac{4GM}{R_1^3 b}(R_1^2- b^2)^{\frac{3}{2}}~
{\rm for}~R < b \leq R_1, \eqno(A2)
$$
which vanishes at $b = R_1$, thereby satisfying the continuity requirement.
Lastly, at $b \leq R$  the clump itself deflects rays inwards
according to Eq. (36).  With the void included, we have
$$
\psi= \frac{4G}{b} \int_0^{\pi/2} m(r) {\rm cos} \alpha d\alpha -
\frac{4GM}{b} 
\left[ 1 - \left(1 - \frac{b^2}{R_1^2} \right)^{\frac{3}{2}} \right], \eqno(A3)
$$
where $r = b{\rm sec} \alpha$ and $m(r)$ is the mass within radius $r$
($m(R) = M$).  
Again, the two values of $\psi$ at the $b = R$ boundary match.

Next, let the source be at an infinite distance away, and the clump-void
system be centered at distance $x$ from the observer.  The angular
magnification formula of Eq. (10) reduces, in the present context, to
$$
\eta = \frac{x}{2} \left(\frac{\psi}{b} + \frac{d\psi}{db} \right) \eqno(A4)
$$
Note that when
light transits the void region (i.e. $R < b \leq R_1$) there is
demagnification - this is the Dyer-Roeder effect.
The average  weak lensing
magnification is obtained by integrating $\eta$ w.r.t. the
probability element appropriate to a random distribution of clump-void
systems of density $n_0$, viz. $dP = 2 \pi n_0 b d b dx$.  If
the clumpy region spans the range $x=0$ and $x=L$,
$$
\langle\eta\rangle = 4 \pi G n_0 \int_0^L xdx \left[ 
\int_0^{\infty} bdb \int_b^{\infty}
\frac{4\pi r \rho(r)dr}{\sqrt{r^2 - b^2}} - \frac{3M}{R_1^2}
\int_0^{R_1} 
\left(1 - \frac{b^2}{R_1^2} \right)^{\frac{1}{2}} b d b \right] = 0, \eqno(A5)
$$
where Eq. (39) was used to calculate the first term in the
square parentheses.
Thus, as long as
the light signals pass {\it through} a sufficient number of clumps and
voids, the average 
source size does not deviate from the Euclidean benchmark.

\vspace{3mm}

\noindent
{\bf Appendix B - on how to average the magnification if strong
lensing is included}

To establish the correct averaging procedure it is only necessary to
consider one foreground clump, placed at the center of the field of
a large, circular, and uniformly illuminated background source.
Take a  small annulus of emission, the undeflected and actual light paths
connecting it and the observer
pass by the clump at impact parameter $b_0$ and $b$ respectively.
In the weak lensing limit,
it is unnecessary to distinguish $b$ from $b_0$.  In the
absence of the lens, the
solid angle subtended by the annulus at the observer is
$$
d \tilde{\omega} = (1+z_l)^2 \frac{2\pi bdb}{x_l^2}.
$$
In the presence of the lens $d \tilde{\omega}$
is increased by the amount $4\pi (1+z_l)^2 \eta bdb/x_l^2$ where, according
to section 8, $\eta$ is given by
$$
\eta = \frac{L}{2}\left(\frac{\psi}{b}
+ \frac{d \psi}{db}\right) =
2GL \int_b^{\infty} \frac{4\pi r \rho(r) dr}{\sqrt{r^2 - b^2}}, \eqno(B1)
$$
with $L = x_l (x_s - x_l)/[(1+z_l) x_s]$.  

If the unlensed source  occupies (in
projection on the lensing plane) a circular area of physical radius $B_0$
as measured at $z=z_l$, the {\it average} percentage magnification of its
observed solid angle may be expressed as 
$$
\frac{\delta \tilde{\omega}}{\tilde{\omega}} = 
\frac{1}{\pi B_0^2} \times 8 \pi GL \int_0^\infty bdb
\int_b^{\infty} \frac{4\pi r \rho(r) dr}{\sqrt{r^2 - b^2}} =
\frac{8GML}{B_0^2} = 2 \langle\eta\rangle,
\eqno(B2)
$$
where in the second last equality used was made of Eq. (39), and
in the last equality the
average of $\eta$ is defined as
$$
\langle\eta\rangle = \frac{\int_0^B 2\pi \eta bdb}{\int_0^B 2\pi bdb},
\eqno(B3)
$$
and matches Eqs. (14) and (15) of the main text with
$1/\int 2\pi bdb$ replacing $n(z) \delta x_l/(1+z_l)$, since in
this Appendix the
number of clumps within a comoving slice at distances $x_l \rightarrow x_l
+ \delta x_l$ is exactly one.
Note that from Eq. (B2) the percentage magnification depends
only on the mass of the enclosed clump - it is not affected by the
details of the clump's internal density profile.  This is reasonable,
because all forms of lensing conserve surface brightness (i.e. no
overlap of emission regions in the image possible)
the extra solid angle the magnified image claims is simply given
by the amount of outward deflection of the `boundary rays' of the source -
a process which relates only to the clump mass - because
such rays are too far away
from the clump to be mindful of its inner structure.  

On the other
hand, we know that the above consideration is limited to the
regime of weak lensing, so the question is whether inclusion of
strong lensing modification of the integrand
in Eq. (B1), which is inevitable as one
reaches the bottom end of the integration,
would lead to new terms carrying
extra parameter dependence to jeopardize our hitherto
consistent result.

To see why the answer is no, we must now extend the treatment to accomodate
the possibility of strong lensing, i.e. the relationship
$b_0 \approx b$ is no longer rigorous enough.  Rather, it has to
be expressed more precisely as
$$
b_0 = b - L \psi(b). \eqno(B4)
$$
We may define the {\it reciprocal magnification} of the annulus area by
the formula
$$
J = \frac{1}{\mu} = \frac{b_0 db_0}{bdb} \eqno(B5)
$$
We see that while the average {\it direct} area
magnification $\mu$ 
$$
\langle\mu\rangle = \frac{\int_0^B 2\pi \mu bdb}{\int_0^B 2\pi bdb}, \eqno(B6)
$$
with  asymptotic boundary parameters related by
$$
B_0 = B - L \psi(B) = B - \frac{4GML}{B}, \eqno(B7)
$$
is divergent because at sufficient low $b$ one encounters the caustic
$J = 0$ which is clearly not a weak lensing 
phenomenon, the average of $J = 1/\mu$ is free from
infinity problems.  Thus our undertaking to evaluate the average
reciprocal magnification is not motivated by the need to obtain
the `desired' outcome, but to obtain a meaningful outcome.  Explicitly
$$
\langle J \rangle = \left\langle \frac{1}{\mu} \right\rangle 
= \frac{\int_0^B 2\pi J bdb}
{\int_0^B 2\pi bdb} = \frac{B_0^2}{B^2}, \eqno(B8)
$$
where in reaching the last expression use was made of the fact that
$J$ is the Jacobian of transformation from $bdb$ to $b_0 d b_0$.
Note also from Eqs. (B3), (B6), and (B8) that the calculation of 
all the averages are consistent.
  
The average percentage magnification of the image area, including the
effect of strong lensing of the central rays, is now  equal to
$\langle J \rangle^{-1} - 1 = (B^2 - B_0^2)/B_0^2$.  By Eq. (B7), this
is just $8GML/B_0^2$, in agreement with Eq. (B2).  
We succeeded in proving that the formula for $\langle\eta\rangle$ in
Eq. (B3), which as explained in the material immediately following this
equation reflects exactly the same
averaging procedure adopted in the main text of
the entire paper, is appropriate to the full regime of weak and strong
lensing.  Therefore, {\it the statistical cancellation between lensing
magnification and the Dyer-Roeder beam, sections 4 and 8, is a robust
conclusion not contingent upon the strength of the lensing.}  Our Eq. (40) for
the variance $(\delta \eta)^2$ is, for this reason, also
valid at arbitrarily low impact parameters.

Those interested in `seeing beyond the mathematics' to actually understand
how the numerator integral
for $\langle J \rangle$ in Eq. (B8) yields $\pi B_0^2$ despite the
presence of multiple images in strong lensing should  consider 
in detail how the
integration is carried out in $b_0$ space  Specifically,
in Eq. (B4)
$b_0$ does not change monotonically with $b$.  As the latter decreases there
comes a place, say $b=b_1$, at which the corresponding $b_0 =
b_{10} =0$.  
After this, initially for $b < b_1$, $b_0$ turns negative.  For any
non-singular central density function $\rho(r)$, however,
$\psi(b) \rightarrow 0$
as $b \rightarrow 0$, so when $b$ decreases further from $b=b_1$ towards
$b=0$ a point $b=b_2$ will come at which $b_0$ reaches a minimum
of $b_0=b_{20}$, thereafter increasing again towards $b_0=0$.  Obviously
$J=0$ at both $b=b_1$ and $b=b_2$.  Within the strong lensing regime,
then,  there
are in general three images of any given source pixel, as is well known.  The
first one corresponds to $b > b_1$, and is on the same side of the center
(on-axis position) as the source pixel.  The second and third images
correspond to $b_2 < b < b_1$ and $0 < b < b_2$ respectively, and lie on the
opposite side.   For the first and third image, $J > 0$; for the second,
$J < 0$.  The $J$-integral may therefore be separated into three, viz.
$$
\int_0^B 2\pi J bdb = \int_{b_1}^B 2\pi J bdb + \int_{b_2}^{b_1} 2\pi J bdb
+ \int_0^{b_2} 2\pi J bdb. \eqno(B9)
$$
Changing variable to $y = |b_0|$and applying
Eq. (B5), one may see how the integration
takes place in $b_0$ space, with the sign of $J$ for the three images
assuming particular importance:
$$
\int_0^B 2\pi J bdb = \int_0^{B_0} 2\pi ydy - \int_0^{|b_{20}|} 2\pi ydy
+ \int_0^{|b_{20}|} 2\pi ydy = \pi B_0^2, \eqno(B10)
$$
in agreement with Eq. (B8).

\vspace{3mm}

\noindent
{\bf Appendix C: On the convergence fluctuation of a large source}

Like the problem in Appendix B, the necessary theorem here 
may be established by
considering only one slice of redshift.  Within this slice let us
introduce  
transverse coordinates $\vec y$ and suppose that
for one clump at $\vec y_1$ the value of $\eta$ 
for a small bundle of light rays at $\vec y$
is given by $\eta(\vec y)=f(\vec y-\vec y_1)$.  If the
number of clumps per unit area is ${\tt n}$, the average of
$\eta$ will be
 $$ 
\bar \eta (\vec y) =
{\tt n} \int f(\vec y - \vec y_1) d^2\vec y_1. \eqno(C1)
 $$
If the averaging is performed over a sufficiently large area
$\bar \eta (\vec y) = \langle\eta\rangle$ should become
independent of $\vec y$.
Moreover, the equation
 $$ 
(\delta \eta)^2={\tt n} \int f^2(\vec y) d^2\vec y.  \eqno(C2)
 $$
gives, by section 5, the variance in $\langle\eta\rangle$.
 
It's also useful to define the correlation function
$\xi(\vec y)$ via
 $$ 
(\delta \eta)^2 \xi(\vec y)=
{\tt n} \int f(\vec y')f(\vec y+\vec y')d^2\vec y'. \eqno(C3)
$$
Clearly, $\xi(\vec 0)=1$, and $\xi(\vec y)$ falls off with $y$ on a scale
characteristic of the size of a clump.  We can therefore write
 $$ 
A_1=\int\xi(\vec y)d^2\vec y. \eqno(C4)
 $$
as {\it definition} of the area $A_1$ of a clump.

To calculate the mean and variance in the magnification of
a background source of projected area $A$ at the lensing plane,
let us first consider the effect of all the clumps in a really large area
$S \gg A$.  The probability that
within $S$ there are exactly ${\cal N}$ clumps at the positions
$\vec y_1,\dots,\vec y_{\cal N}$, say, within small areas
$d^2\vec y_1,\dots,d^2\vec y_{\cal N}$, is given by the Poisson
distribution
$$
d^{2{\cal N}} P =
e^{-{\tt n} S} {\tt n}^{\cal N} \prod_{j=1}^{\cal N} d^2\vec y_j.  \eqno(C5)
$$
The total increase in the
area of $A$ due to lensing by these clumps  is $2\eta_A A$ where
$$
\eta_A A=\int_A d^2\vec y\sum_{j=1}^{\cal N} f(\vec y-\vec y_j). \eqno(C6)
$$
To find the average, we integrate (C6) over the measure
(C5), and sum over ${\cal N}$.  The result is
$$
\langle\eta_A\rangle A=\int_A d^2\vec y\,e^{-{\tt n} S} 
\sum_{{\cal N}=1}^\infty 
\frac{{\tt n}^{\cal N}}{{\cal N}!} \int_S d^2\vec y_1\dots d^2\vec y_{\cal N}
\sum_{j=1}^{\cal N} f(\vec y-\vec y_j). \eqno(C7)
$$
Note that the ${\cal N}!$ factor is needed to compensate for multiple counting
of permutations of the clumps.  Also, as already explained in Appendix B,
this averaging procedure has correctly taken into account strong
lensing effects as well.  Thus we obtain
$$
\bar\eta_A=e^{-{\tt n}S}
\sum_{{\cal N}=1}^\infty \frac{{\tt n}^{{\cal N}-1}}{({\cal N}-1)!}
S^{{\cal N}-1} \langle\eta\rangle \left[\frac{1}{A}\int_A d^2\vec y \right]
= \langle\eta\rangle,  \eqno(C8)
$$
where in reaching Eq. (C8) we observe that each
$\int_S d^2 \vec x_i$ term in the summation over $j$ in Eq. (C7)
gives the same contribution: one of the integrals over $\vec x_i$ yields,
by Eq. (C1), 
$\langle\eta\rangle/{\tt n}$, and the others altogether yield $S^{{\cal N}-1}$.
 
Concerning the average of $\eta_A^2$, we now have to
take the square of Eq. (C7) and integrate over the measure of Eq. (C6), i.e.
$$
\langle\eta_A^2\rangle A^2=\int_A d^2\vec y\int_A d^2\vec
y'\,e^{-{\tt n} S}\sum_{{\cal N}=1}^\infty \frac{{\tt n}^{\cal N}}
{{\cal N}!} \int_S d^2
\vec y_1\dots d^2\vec y_{\cal N}\sum_{j,k=1}^{\cal N} f(\vec y-
\vec y_j) f(\vec y'-\vec y_k). \eqno(C9)
$$
In Eq. (C9) it is necessary to separate the terms with $j=k$ from the
others.  All terms with with $j\ne k$ are equal, totalling
$$
e^{-{\tt n}S}\sum_{{\cal N}=2}^\infty
\frac{{\tt n}^{{\cal N}-2}}{({\cal N}-2)!} S^{{\cal N} -2}
\langle\eta\rangle^2  \int_A d^2\vec y \int_A d^2\vec y'
=\langle\eta\rangle^2 A^2, \eqno(C10)
$$
i.e. these terms add up to the very quantity that needs to be
subtracted from $\langle\eta_A^2\rangle A^2$ to obtain the variance.
The variance is then simply given by the sum total of all the $j=k$ terms.
We arrive at
$$
(\delta\eta_A)^2 A^2= e^{-{\tt n} S}\sum_{N=1}^\infty 
\frac{{\tt n}^{{\cal N}-1}}{({\cal N}-1)!} S^{{\cal N} -1}
\int_A d^2\vec y\int_A d^2\vec y'
\int_S d^2\vec y_1~{\tt n}~
f(\vec y- \vec y_1)f(\vec y'-\vec y_1). \eqno(C11)
$$
By Eq. (C3) we have, finally,
$$
(\delta\eta_A)^2 A^2=(\delta\eta)^2 \int_A d^2\vec y\int_A
d^2\vec y'\,\xi(\vec y'- \vec y). \eqno(C12)
$$

Defining $N=A/A_1$ as the ratio of the source area to the clump area,
then, so
long as $N$ is large (i.e. away from the
$N \leq 1$ regime) $\vec y$ will not be too near the
boundary, and the integral over $\vec y'$ will give $A_1$, while
the remaining integral over $\vec y$ is  $= A$. 
Thus, apart from small (boundary-effect) corrections, 
$$
(\delta\eta_A)^2=(\delta\eta)^2 \frac{A_1}{A}=
\frac{(\delta \eta)^2}{N}.  \eqno(C13)
$$

\noindent
{\bf References}

\noindent
Bahcall, N. 1988, ARAA, 26, 631.

\noindent
Barber, A.J., 2000, MNRAS, 318, 195.

\noindent
Barris, B.J. et al 2004, ApJ, 602, 571.

\noindent
Bartelmann, M. \& Schneider, P., 2001, Physics Reports, 340, 291.

\noindent
Bennett, C.L. et al, 2003, ApJS, 148, 1-27.

\noindent
Carlberg, R.G. et al 1997, ApJ, 485, L13.

\noindent
Dalal, N., Holz, D.E., Chen, X., \& Frieman, J.A., 2003, ApJ, 585, L11.

\noindent
Dubinski, J., \& Carlberg, R.G. 1991, ApJ, 378, 496.

\noindent
Dyer, C.C., \& Roeder R.C., 1972, ApJ, 174, L115.


\noindent
Fukugita, M., 2003, in `Dark matter in galaxies', IAU Symp. 220, Sydney
(astro-ph/0312517).

\noindent
Fukugita, M., Hogan, C.J., \& Peebles, P.J.E., 1998, ApJ, 503, 518.

\noindent
Hashimoto, Y., Barcons, X., Boehringer, H., Fabian, A.C., \\
\indent Hasinger, G., Mainieri, V., Brunner, H., 2004, A \& A, 417, 819.

\noindent 
Hashimoto, Y., Hasinger, G., Arnaud, M., Rosati, P., Miyaji, T.,
2002, A \& A, 381, 841.

\noindent
Jones, L.R., McHardy, I., Newsom, A, Mason, K, 2002, MNRAS, 334, 219.

\noindent
Kibble, T.W.B., \& Lieu, R., 2005, ApJ, in press (astro-ph/0412275).

\noindent
Lieu, R., \& Mittaz, J.P.D., 2005, ApJ, 623, L1 (astro-ph/0409048).



\noindent
Maughan, B.J., Jones, L.R., Ebeling, H., Perlman, E., Rosati, P., \\
\indent Frye, C., Mullis, C.R., 2003. ApJ, 587, 589.

\noindent
Mulchaey, J, 2000, ARAA, 38, 289.

\noindent
Navarro, J.F., Frenk, C.S., \& white, S.D.M. 1995, MNRAS, 275, 56.

\noindent
Navarro, J.F., Frenk, C.S., \& white, S.D.M. 1996, ApJ, 462, 563.

\noindent
Navarro, J.F., Frenk, C.S., \& white, S.D.M. 1997, ApJ, 490, 493.



\noindent
Pfrommer, C., 2002, Cosmological weak lensing of the cosmic  microwave
background \\
\indent by large scale structures, Diploma Thesis, Friedrich-Schiller
Universit\"at Jena.

\noindent
Ramella, M., Geller, M.J., Pisani, A., \& da Costa, L.N.,
2002, AJ, 123, 2976.

\noindent
Ramella, M. et al, 1999, A \& A, 342, 1.


\noindent
Tonry, J.L. et al 2003, ApJ, 594, 1.

\noindent
Wambsganss J., Cen, R., Xu, G, \& Ostriker, J.P., 1997, ApJ, 475, L81. 

\noindent
Weinberg, S., 1976, ApJ, 208, L1.

\noindent
Younger, J.D., Bahcall, N.A., Bode, P., 2005, ApJ, 622, 1.

\noindent
Zabludoff, A.I., \& Mulchaey, J.S., 1998, ApJ, 496, 39.

\vspace{2mm}

\noindent
{\it Note added in Proof (dated 14th June, 2005):}
We are now aware of the paper by Scranton et al (ApJ in press,
astro-ph/0504510), which reported the first secure detection of magnification
dispersion, or convergence fluctuation, among a large sample
of distant quasars.  This provides the necessary observational evidence that
the effect discussed in our present work is real, and further
elevates the profile of
the question on why the CMB exhibits no sign of having been lensed.

\end{document}